\documentstyle[epsfig,aps]{revtex}
\begin{document}
\twocolumn[\hsize\textwidth\columnwidth\hsize
           \csname @twocolumnfalse\endcsname]
\draft
%%%%%%%%%%%%%%%%%%%%%%%%%%%%%%%%%%%%%%%%%%%%%%%%%%%%%%%%%%%%%%%%%%%
\title{Off-shell $\omega$ production in proton-proton collisions 
near threshold}
\author{C. Fuchs$^a$, M.I. Krivoruchenko$^{a,b}$, 
H.L. Yadav$^{a,c}$, Amand Faessler$^a$, B.V. Martemyanov$^{a,b}$, 
K. Shekhter$^a$}
\address{$^a$ Institut f\"ur Theoretische Physik der Universit\"at T\"ubingen, 
Auf der Morgenstelle 14, D-72076 T\"ubingen, Germany}
\address{$^b$ Institute for Theoretical and Experimental
Physics, B. Cheremushkinskaya 25, 117259 Moscow, Russia}
\address{$^c$ Physics Department, Rajasthan University, 
Jaipur-302004, India}
\maketitle  
%%%%%%%%%%%%%%%%%%%%%%%%%%%%%%%%%%%%%%%%%%%%%%%%%%%%%%%%%%%%%%%%%%%
\begin{abstract}
The $\omega$ production in nucleon-nucleon collisions is described through 
decays of intermediate nucleon resonances. The near-threshold 
$pp\rightarrow pp\omega$ cross section is found to be dominated 
by the off-shell production of $\omega$ mesons with masses far below 
the physical $\omega$ mass. The $N^*(1535)$ resonance plays 
thereby a crucial role. Due to a strong $N\omega$ coupling, this resonance 
leads to an off-shell contribution which is about one order of magnitude 
larger than the experimentally measured contribution from the $\omega$ peak. 
After a subtraction of the theoretical "background" from the 
off-shell $\omega$ 
production, the available data are accurately reproduced over the entire
energy range from 5 MeV up to several GeV above the threshold.
The scenario of a weaker $N^*(1535)\rightarrow N\omega $ decay mode 
which is still consistent with electro- and photoproduction data 
is discussed as well. In the latter case, the off-shell contribution
to the  $pp\rightarrow pp\omega$ cross section is substantially 
reduced but the description of the experimental cross section is poor 
above threshold.
\end{abstract}

\pacs{13.60.Le, 14.20.Gk, 25.40-h}
%\newpage
%%%%%%%%%%%%%%%%%%%%%%%%%%%%%%%%%%%%%%%%%%%%%%%%%%%%%%%%%%%%%%%%%%%
\section{Introduction}
%%%%%%%%%%%%%%%%%%%%%%%%%%%%%%%%%%%%%%%%%%%%%%%%%%%%%%%%%%%%%%%%%%%
The studies of vector meson production in nucleon-nucleon collisions 
are motivated by several facts. First, the short range part of the nucleon-nucleon 
($NN$) interaction is dominated by the isoscalar $\omega$ meson exchange 
\cite{bonn} and thus information from inelastic $NN$ collisions can 
contribute to better understanding of the nuclear forces. 
In a dense hadron environment which exists in cores of massive neutron stars 
or is transiently created in energetic heavy-ion reactions 
one can expect significant changes of the 
quark condensates, which manifest themselves in mass shifts of the vector 
mesons \cite{BR}, suggested also by QCD sum rules \cite{QCDSR}. 
In a hadronic picture the coupling to many-body correlations modifies 
the in-medium masses and the spectral properties of the 
mesons \cite{width,friman97}. To search for modifications
of the in-medium 
properties of vector mesons is the major issue of dilepton spectroscopy 
in energetic heavy-ion collisions \cite{dilep}.  

However, for the study of medium effects in heavy-ion reactions the 
theoretical understanding of vector meson production in 
elementary processes is a prerequisite. In the vicinity of the 
threshold only the $\omega$ 
production has been studied experimentally. This is due to the small 
$\omega$ width of $8.4$ MeV which allows an identification of the 
narrow $\omega$ peak in missing mass spectra already close to 
threshold. Data for the $\omega$ production in proton-proton 
($pp\rightarrow pp\omega$) collisions at small excess energies 
were recently taken at SATURNE \cite{hibou99} and with the COSY-TOF 
spectrometer \cite{cosy01}. The excess energy $\epsilon$ is thereby defined 
as $\epsilon = \sqrt{s} - (2m_p + m_\omega )$ with $\sqrt{s}$ being 
the proton-proton center-of-mass energy, $m_p$ the proton mass and 
$m_\omega = 782$ MeV the physical $\omega$ mass at its pole value. The SATURNE 
experiment was performed very close to threshold ($ \epsilon = 4\div 30$ MeV) 
whereas the COSY-TOF Collaboration measured the $\omega$ production at 
$ \epsilon =92$ and 173 MeV. Both experiments served also to examine 
the validity of the Okubo-Zweig-Iizuka (OZI) selection rule \cite{OZI}. 
As compared to the  $\omega$ cross section, the $\phi$ meson which couples to the 
strangeness content of the nucleon should be strongly suppressed by the 
OZI rule which forbids disconnected quark line diagrams. 
The assumption of only a small OZI violation leads to a 
$\phi$ over $\omega$ ratio of $\sim 4\cdot 10^{-3}$ 
in $pp$ reactions at comparable excess energies \cite{OZI2} while 
experimentally this ratio was found to be about one order of magnitude larger 
\cite{hibou99,cosy01,disto}.

Microscopic calculations for the $\omega$ production in 
nucleon-nucleon collisions are, however, rare. In ref. \cite{Nak98} 
the on-shell $\omega$ production was described through the 
coupling to nucleon currents and to meson exchange currents 
including thereby the $NN$ final-state interaction (FSI).
The $\omega$-meson was treated as an elementary field and thus 
off-shell effects in the $\omega$ production were missing from the beginning.
Adjusting the relevant form factors, the experimental data \cite{hibou99}
in the vicinity of the threshold were well reproduced. 

In the present approach the vector meson production is described through 
a two-step mechanism via the excitation of nuclear resonances.  
This picture is motivated by existence of a variety of nucleon resonances 
which have large branchings to the $N\rho$ decay modes, such as
$N^*(1520)\frac{3}{2}^-,~N^*(1535)\frac{1}{2}^-$, and $N^*(1720)\frac{3}{2}^+$.
The couplings to the $\omega$-mesons have been observed in 
multichannel $\pi N$ partial wave analyses \cite{Vrana:1999nt,Manley:1992yb} 
and were predicted by quark models \cite{Koniuk:1982ej}.
A strong coupling of the $N\omega$ system to nuclear resonances has also been 
reported in \cite{lutz02} using an EFT coupled channel approach. 
It is further a well established fact 
that within a hadronic picture the 
coupling to nuclear resonances is a major source for the 
modification of vector meson properties in a dense nuclear environment. 
E.g. in the medium the $\rho$ acquires its major modifications 
by the coupling to $N^*$-hole excitations, especially the 
$N^*(1520)N^{-1}$ \cite{width,friman97}. 

The resonance model provides a unified description of a large variety 
of phenomena such as nucleon electro- and photoproduction, mesonic 
decays of nuclear resonances and vector meson production in elementary 
nucleon-nucleon reactions. The resonance decay modes to vector mesons 
can be fixed from electro-, photoproduction and mesonic 
decay data \cite{friman97,post01,krivo02} 
and the resonance production cross sections from $NN\rightarrow NN\pi$ 
scattering data \cite{Teis}. In this way, no new parameters enter
into the description of the vector meson production in nucleon-nucleon 
reactions. The model is described in the next section. Close 
to threshold the behavior of the $pp \rightarrow pp\omega$ cross section 
turns thereby out to depend in a crucial way on properties of the 
$N^*(1535)$ resonance. The analysis of \cite{krivo02} on which the 
present calculations are based predicts a large $N^*(1535)N\omega$ 
coupling which, as will be seen in the following, gives rise to large 
off-shell contributions in the $pp \rightarrow pp\omega$ cross section 
close the $\omega$ production threshold. To estimate the influence of the 
$N^*(1535)$ we consider also the case of a weaker  $N^*(1535)N\omega$ 
coupling which is still compatible with the existing 
electro-, photoproduction and mesonic decay data for the $N^*(1535)$. 
When the $pp \rightarrow pp\omega$ cross section is compared to 
experimental data one has to keep in mind that experimentally only the 
distinct $\omega$ peak can be resolved in missing mass spectra. Off-shell 
contributions are attributed to the general experimental background. 
In particular when off-shell 
contributions are large this fact has to be taken into account and
only the experimentally measurable part of the cross section should 
be considered. We apply such a procedure when the 
comparison to data is performed in Sec. III. The experimental data, 
in particular in the vicinity of the threshold, support the picture 
of a strong coupling of the  $N^*(1535)$ to the $N\omega$ channel 
with the consequence of large off-shell contributions to the cross 
section.  
%%%%%%%%%%%%%%%%%%%%%%%%%%%%%%%%%%%%%%%%%%%%%%%%%%%%%%%%%%%%%%%%%%%
\section{Resonance model}
%%%%%%%%%%%%%%%%%%%%%%%%%%%%%%%%%%%%%%%%%%%%%%%%%%%%%%%%%%%%%%%%%%%
The vector meson production is described through 
a two-step mechanism via the excitation of nuclear resonances, i.e. 
$NN \rightarrow NR$, $R \rightarrow NV$, with $V = \rho$, $\omega$. 
The nucleon-nucleon cross sections for the resonance 
production $NN \rightarrow NR$ have been determined 
by fitting the available data on pion, double-pion, $\eta$ and $\rho$ 
production \cite{Teis}. The reactions $NN \rightarrow NNP, NNV$, 
and $NNPP$ ($P= \pi, \eta, ...$) are described by the two-step process 
$NN \rightarrow NR$ with the subsequent decays of the 
resonances $R$'s \cite{Teis}. The influence of the $NN$ FSI which 
is generally expected to be 
important near the thresholds is thereby effectively included in 
the phenomenological matrix elements ${\cal M}_R$'s 
for the resonance production. On the other hand, the present approach 
accounts also effectively for the FSI between 
one nucleon and the produced vector meson. To 
take the FSI between the nucleon and the meson 
in specific partial waves into account is equivalent 
to include those nucleon resonances into the reaction dynamics 
which have the same quantum numbers as the corresponding 
partial waves (final-state interaction theorem \cite{watson}). 
The nucleon resonance propagators, in particular, reoccur due to the 
division of the bare amplitudes by Jost functions evaluated 
for $NV$ resonating phase shifts.

In refs. \cite{friman97,post01,krivo02}, the Vector Meson Dominance 
(VMD) model was applied to determine 
the $RN\omega$ coupling strengths from radiative and mesonic 
decay data. In its monopole form, i.e. only including the ground-state 
$\rho$ and $\omega$, the naive VMD is known to underestimate 
systematically the mesonic $R\rightarrow N\rho$ decays when a normalization to  
radioactive $R\rightarrow N\gamma$ branchings is performed 
\cite{Faessler:2000md}.
The inclusion of higher radial $\rho$ meson excitations in the VMD helps 
to resolve this problem
\cite{krivo02,Faessler:2000md} and provides the correct asymptotes of the 
$R\rightarrow N\gamma$ transition form factors given by the quark counting 
rules \cite{Matveev:1973ra}. The extended (e)VMD model assumes that 
radial excitations $\rho (1450), \rho (1700),...$ 
interfere with the ground-state $\rho$-mesons in radiative processes. 
Already in the case of the nucleon form factors, radially excited vector
mesons should be added in order to provide a dipole behavior for the Sachs
form factors and describe the experimental data \cite{Hohler:1976ax}. 
Details of the calculations of the  magnetic, electric, 
and Coulomb $R\rightarrow N\gamma$ transition form factors and the 
branching ratios of the nucleon resonances 
can be found in refs. \cite{krivo02,Krivoruchenko:2001hs}. 
The model parameters are fixed 
from photo- and electro-production data and using results
from $\pi N$ scattering multichannel partial-wave analyses. Where
experimental data are not available, predictions from non-relativistic
quark models are used. In \cite{Faessler:1999de} the eVMD model was 
successfully applied to a systematic study of meson decays to 
dilepton pairs and in \cite{Faessler:2000md} to the description of dilepton 
production in $pp$ reactions \cite{DLS}.

\begin{figure}[h]
\begin{center}
\leavevmode
\epsfxsize = 7cm
\epsffile[20 20 520 440]{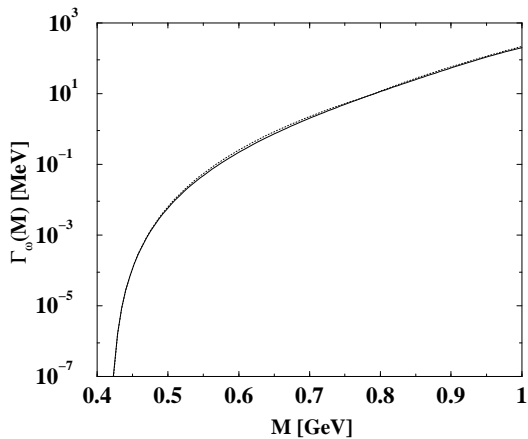}
\end{center}
\caption{The $\omega$-meson decay width $\Gamma_{\omega}(M)$ 
versus the off-shell $\omega$-meson mass $M$. 
The solid line shows the calculation (\ref{GM}) according to 
Gell-Mann, Sharp, and Wagner \protect\cite{gom}.
The dotted line is the parametrisation of Eq. (\ref{param}).
}
\label{Gam_fig}
\end{figure}

The $pp\rightarrow pp\omega$ cross section is given as follows
\begin{eqnarray}
&&\frac{d\sigma (s,M)^{pp\rightarrow pp\omega}}{dM^{2}}
=
\nonumber \\
&&\sum_{R}\int_{(m_{p}+M)^{2}}^{(\sqrt{s}-m_{p})^{2}}d\mu ^{2}
\frac{ d\sigma (s,\mu )^{pp\rightarrow pR}}{d\mu ^{2}}
\frac{dB(\mu,M)^{R\rightarrow p\omega }}{d M^{2}}~~.
\label{sigNNV}
\end{eqnarray}
The cross sections for the resonance production are given by 
\begin{equation}
%d\sigma (s,\mu )^{pp\rightarrow pR} = 
%\frac{|{\cal M}_R|^2 ~ p^*(\sqrt {s},\mu,m_{p})}{16 p_i s\pi^2} 
%\frac{\mu \Gamma_{\rm tot}^R (\mu) d\mu^2 }
%{(\mu^2-m_{R}^2)^2 +(\mu\Gamma_{\rm tot}^R(\mu))^2},
d\sigma (s,\mu )^{pp\rightarrow pR} = 
\frac{|{\cal M}_R|^2}{16 p_i \sqrt s \pi^2}\Phi_2(\sqrt {s},\mu,m_{p})dW_R(\mu)
\label{sigNR}
\end{equation}
with $\Phi_2(\sqrt {s},\mu,m_{p}) = \pi p^*(\sqrt {s},\mu,m_{p})/{\sqrt s}$ 
being the two-body phase space, 
$p^*(\sqrt {s},\mu,m_{p})$ the final c.m. momentum, $p_i$ the 
initial c.m. momentum, and $\mu$ and $m_R$ the running and pole masses 
of the resonances, respectively. $m_p$ is the proton mass. The
mass distribution $dW_{R}(\mu)$ of the resonances is described by the 
standard Breit-Wigner formula:
\begin{equation}
dW_{R}(\mu) = \frac{1}{\pi}\frac{\mu \Gamma_R (\mu)d\mu^2 }
{(\mu^2-m_{R}^2)^2 +(\mu\Gamma_R (\mu))^2}~.
\label{BW}
\end{equation}
The sum in (\ref{sigNNV}) runs over the nucleon resonances given in Table 1. 
This includes all well established (4$*$) resonances quoted by the PDG \cite{pdg}. 
The branching to the $\omega$ decay mode is given by 
\begin{equation}
dB(\mu,M)^{R\rightarrow p\omega } = 
\frac{\Gamma_{\rm N\omega}^R (\mu,M)}{\Gamma_R (\mu)}dW_{\omega}(M)~~.
\label{bra}
\end{equation}
The $\omega$ mass distribution $dW_{\omega}(M)$ is also described 
by a Breit-Wigner distribution, i.e. substituting $R \rightarrow \omega$ 
and $\mu \rightarrow M$ in Eq. (\ref{BW}). 
The energy dependence of the $\omega$ width 
$\Gamma_\omega (M)$ can be calculated according to the two-step 
process $\omega\rightarrow \rho\pi\rightarrow 3\pi$, 
as proposed by Gell-Mann, Sharp, and Wagner \cite{gom}.
The effective vertices describing the $\omega \rightarrow \rho \pi $ and $%
\rho \rightarrow 2\pi $ decays have the form ${\cal L}_{\omega \rho \pi
}=f_{\omega \rho \pi }\epsilon _{\tau \sigma \mu \nu }\partial _{\sigma
}\omega _{\tau }\partial_\nu \rho^{\alpha }_\mu \pi^{\alpha }$ 
and ${\cal L}_{\rho \pi \pi
}= - \frac{1}{2}f_{\rho \pi \pi }\epsilon _{\alpha \beta \gamma }\rho _{\mu }^{\alpha }\pi
^{\beta }\partial _{\mu }\pi ^{\gamma },$ with $f_{\omega \rho \pi }=16.3$ GeV$%
^{-1}$ and $f_{\rho \pi \pi }=6.03$ (see e.g. [17]). The decay width can be found to be
\begin{eqnarray}
\Gamma _{\omega }(M) &=&\frac{1}{48\pi ^{5}M}f_{\omega \rho \pi }^{2}f_{\rho
\pi \pi }^{2}\int_{4m_{\pi} ^{2}}^{(M-m_{\pi} )^{2}}dM^{\prime 2}\int_{-1}^{+1}\frac{%
dz}{2}
\nonumber \\
&&|D(k^{+}+k^{-}) +D(k^{+}+k)+D(k^{-}+k)|^{2}
\nonumber \\
&&\times(m_{\pi}^{2}p^{2}q^{2}-p^{2}(kq)^{2}-q^{2}(kp)^{2})  
\nonumber \\
&&\times \Phi _{2}(M,M^{\prime },m_{\pi} )\Phi _{2}(M^{\prime },m_{\pi} ,m_{\pi} )
\label{GM}
\end{eqnarray}
where $m_{\pi}$ is the pion mass, 
$k^{+},$ $k^{-},$ and $k$ are the pion $\pi ^{+},$ $\pi ^{-},$ and $%
\pi ^{0}$ momenta, $p=k^{+}+k^{-},$ $q=(k^{+}-k^{-})/2,$ $p^{2}=M^{\prime 2},
$ $kp=(M^{2}-M^{\prime 2}-m_{\pi} ^{2})/2,$ $q^{2}=-p^{*2}(M^{\prime },m_{\pi} ,m_{\pi} ),$
$kq=-p^{*}(M,M^{\prime },m_{\pi} )p^{*}(M^{\prime },m_{\pi} ,m_{\pi} )Mz/M^{\prime }$. 
The value of $z$ is the cosine between the 
vectors {\bf q} and {\bf k} in the c.m. frame of the charged pions. 
$D(p)$ is the $\rho$-meson propagator with the total width of $150$ MeV. The $\omega$ 
width can be parameterized by 
\begin{equation}
\Gamma _{\omega }(M) = \Gamma _{\omega }(m_{\omega})
\left(\frac{p^{*}(M,m_{\pi},2m_{\pi})}{p^{*}(m_{\omega},m_{\pi},2m_{\pi})}
\right)^a
\label{param}
\end{equation}
with the coefficient $a = 8$ at $M<m_{\omega}$ and $a = 10$ at $M > m_{\omega}$. 
Expression (\ref{param}) is compared to Eq. (\ref{GM}) in Fig. \ref{Gam_fig}. 
The $\omega$ width goes to zero at the $3m_\pi$ threshold which is 
also the physical threshold for the $pp\rightarrow pp\omega$ reaction. 
%%%%%%%%%%%%%%%%%%%%%%%%%%%%%%%%%%%%%%%%%%%%%%%%%%%%%%%%%%%%%%%%%%%%%%%%%
%%%%%%%%%%%%%%%%%%%%%%%%%%%%%%%%%%%%%%%%%%%%%%%%%%%%%%%%%%%%%%%%%%%%%%

Nucleon resonances with spin $J>1/2$ 
and arbitrary parity have three independent transition amplitudes, while 
spin-1/2 resonances have only two independent amplitudes. In terms of
the magnetic, electric, and Coulomb couplings $g_M^{(\pm)}$, $g_E^{(\pm)}$, 
and $g_C^{(\pm)}$, the differential
decay widths of nucleon resonances with spin 
$J = l + 1/2$ into an $\omega$-meson with arbitrary mass $M$ 
has the form \cite{krivo02}
\begin{eqnarray}
&&d\Gamma_{N\omega}^R(\mu,M) =\frac{9}{64\pi}\frac{%
(l!)^{2}}{2^{l}(2l+1)!}\times 
\nonumber \\
&&\frac{m_{\pm }^{2}(m_{\mp }^{2}-M^{2})^{l+1/2}(m_{\pm
}^{2}-M^{2})^{l-1/2}}{\mu^{2l+1}m^{2}}  
\left( \frac{l+1}{l}\left| g_{M/E}^{(\pm )}\right| ^{2}\right.
\nonumber \\
&&\left.
+(l+1)(l+2)\left|
g_{E/M}^{(\pm )}\right| ^{2}+\frac{M^{2}}{\mu^{2}}\left| g_{C}^{(\pm
)}\right| ^{2}\right)dW_{\omega}(M),  
\label{GAMMA_l}
\end{eqnarray}
with $m_{\pm} = \mu \pm m_p$. The signs $\pm$ refer to the natural parity ($1/2^-,
3/2^+, 5/2^-,$ ...) and abnormal parity ($1/2^+,
3/2^-, 5/2^+,$ ...). $g_{M/E}^{\pm}$ means $g_{M}^{+}$ or $g_{E}^{-}$. 
The above equation is valid
for $l>0$. For $l=0$ ($J=1/2$) one obtains 
\begin{eqnarray}
d\Gamma_{N\omega}^R(\mu,M) =&&\frac{1 }{32\pi\mu}%
(m_{\pm }^{2}-M^{2})^{3/2}(m_{\mp }^{2}-M^{2})^{1/2}  \nonumber \\
&&\left( 2\left| g_{E/M}^{(\pm )}\right| ^{2}+\frac{M^{2}}{\mu^{2}}\left|
g_{C}^{(\pm )}\right| ^{2}\right)dW_{\omega}(M).  \label{GAMMA_0}
\end{eqnarray}
Due to the subthreshold character of the $\omega$ production in decays of 
on-shell nucleon resonances, the $M$-dependence of the coupling constants 
$g_M^{(\pm)}$, $g_E^{(\pm)}$, and $g_C^{(\pm)}$ can be important. At 
the $\omega$ pole mass $m_{\omega}$ these couplings are proportional to residues of 
the magnetic, electric, and Coulomb transition form factors. The 
corresponding values at $M^2=m_{\omega}^2$ can be found in Table 1.  
We assume that the coupling constants which enter into the covariant 
representation of the form factors are not mass dependent. 
The $M$-dependence of $g_M^{(\pm)}$, $g_E^{(\pm)}$, 
and $g_C^{(\pm)}$ arises then exclusively from the $M$-dependent 
transformation from the covariant basis to the multi-pole basis according to
%%%%%%%%%%%%%%%%%%5
\begin{equation}
g_{T}^{(\pm)}(M^{2})=\sum_{kT^{\prime }}{\rm M}_{Tk}(M^{2}){\rm M}_{kT^{\prime
}}^{-1}(m_{\omega}^{2})g_{T^{\prime }}^{(\pm)}(m_{\omega}^{2}),  \label{RUN}
\end{equation}
with $T,T^{\prime}=M,E,C$. The matrices ${\rm M}_{kT}(M^{2})$ which transform the multi-pole
form factors to the covariant form factors can be found in ref. \cite{krivo02}. 
%%%%%%%%%%%%%%%%%%

According to our analysis of ref. \cite{krivo02} the $N^*(1535)$ is the 
only resonance with a large branching to the $N\omega$ 
channel among those resonances which lie far below the $\omega$ peak. 
Since it turns out that the $\omega$ production reacts very
sensitive to this fact, in particular close to threshold, 
we consider also an alternative scenario with a weaker 
$N^*(1535)N\omega$ coupling. In the eVMD model the 
$N\rho$ and $N\omega$ decays are fixed simultaneously. However, 
in the case of the $N^*(1535)$ there were neither experimental 
data nor quark model predictions for the $N\omega$ channel 
available \cite{krivo02}. The available data in the $N\rho$ 
channel and the fact that quark models provide here partially contradictory 
results leave some freedom in the determination of the $N^*(1535)N\omega$ 
coupling strength. In Fig. \ref{Nfit1_fig} shown in the Appendix,
 we discuss two different fits to the electro- and photoproduction data, 
the $\pi N$ multichannel scattering analyses and the quark model predictions in
more detail. The essential distinction between these two 
procedures lies in the different normalizations 
to the $\rho$-meson decay amplitudes. Koniuk \cite{Koniuk:1982ej}, 
Manley and Saleski \cite{Manley:1992yb} and the PDG \cite{Groom:2000in} 
give similar predictions for the $s_{1/2}$ wave 
and predict the same sign for the $d_{3/2}$ wave. 
The second set of amplitudes stems from quark model calculations 
by Capstick and Roberts \cite{Capstick:1994kb}. Their values 
are noticeable smaller than those proposed in 
\cite{Koniuk:1982ej,Manley:1992yb,Groom:2000in}. Both sets do 
not provide $\omega$-meson amplitudes. To investigate the stability 
of our results, we introduced one point for the 
$\omega$-meson $s_{1/2}$ wave around zero. The original solution \cite{krivo02}, 
which uses the results of refs.\cite{Koniuk:1982ej,Manley:1992yb,Groom:2000in} 
turns out to be rather stable and does not allow for a significant reduction
of the $s_{1/2}$ amplitude. The second set allows the reduction of the 
amplitude by a factor of 6 to 8, however, at the expense of a moderately 
higher $\chi^2$. A further going reduction of the amplitude is hardly possible. 
In the second fit the deviation from the experimental $p^*(1535)$ Coulomb 
amplitude is larger and the reproduction of the transversal $p^*(1535)$ 
amplitude is also worse but the $\rho$-meson amplitudes are
better reproduced than with the first set of input parameters. 
Notice that the $d_{3/2}$ amplitude of the $\omega$-meson was set 
equal to zero to ensure an unique eVMD solution for the fitted data. 
The value of $d_{3/2}$ does not significantly affect other observables. 

In Table 1 the corresponding values for the $N\omega$ decay widths 
are given at the resonance pole masses. The branching of the $N^*(1440)$ 
at the resonance pole is small since it lies deeply below the $N\omega$ 
threshold. However, at higher masses where the $N^*(1440)$ is off-shell 
it receives a sizable $\omega$ decay width. 
In Table 1, the contributions from different partial waves, including relative signs, 
are also shown. In the case of the $N^*(1535)$ the values 
for the two different parameter sets resulting 
in a strong (s), respectively a weaker (w) coupling to the $N\omega$ channel 
are given. The corresponding matrix elements ${\cal M}_R$ for the resonance 
production in $pp$ collisions, Eq. (\ref{sigNR}), are taken from ref. \cite{Teis}.

The energy dependent total resonance width 
$\Gamma_{R}^{[0]} (\mu)$ is scaled according to the $\pi N$ 
phase space and the Blatt-Weisskopf 
suppression factor. In a consistent treatment, the $N\rho$ and $N\omega$ 
decay channels have to be taken into account in the total width
\begin{equation}
\Gamma_R (\mu) = \Gamma_{R}^{[0]} (\mu) +  \Gamma^{R}_{N\rho} (\mu) 
+ \Gamma^{R}_{N\omega}(\mu) +\delta\Gamma_R
\label{gres}
\end{equation}
where $\delta\Gamma_R = - \Gamma^{R}_{N\rho} (m_R) - \Gamma^{R}_{N\omega}(m_R)$ ensures 
the normalization of the total width at the resonance pole mass $m_R$. Since the 
total width appears in the numerator and squared in the denominator of the 
corresponding Breit-Wigner distribution (\ref{BW}) the partial widths 
are not simply added in a perturbative way but the coupled channel 
problem is more complicated. The appearance of new channels shifts 
generally strength from the pole to the tails of the distribution. 
However, the shape of the spectral function around the $\omega$ pole depends 
crucially on the magnitude of the partial decay widths. In the case 
of a strong $N\omega$ coupling, the rapidly increasing partial width 
leads to a strong reduction at 
the $\omega$ threshold, an effective enhancement below the $\omega$ peak, 
and shifts strength further out. A weak coupling, on 
the other hand, leads to a visible $\omega$-peak in the resonance spectral 
function. The consequences of these two different scenarios will
become more clear when the cross sections are considered in the next 
section.

\begin{figure}[h]
\begin{center}
\leavevmode
\epsfxsize = 8cm
\epsffile[70 40 480 430]{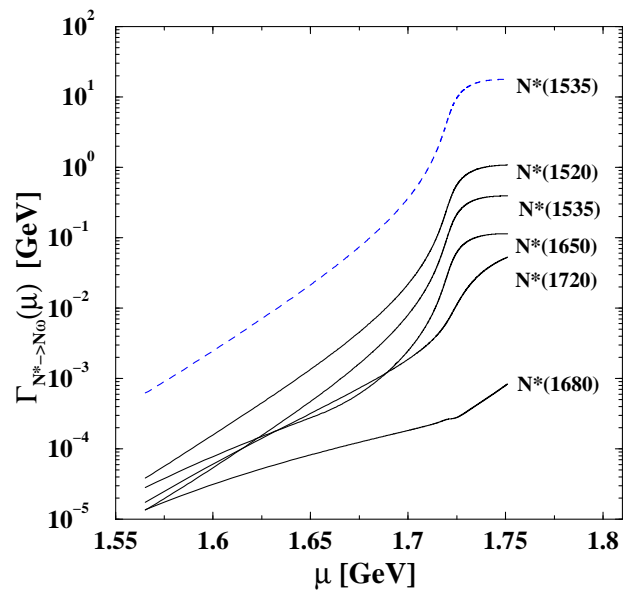}
\end{center}
\caption{Energy dependence of the $N^* \rightarrow N\omega$ 
widths versus the off-shell masses $\mu$ of the resonances. 
For the $N^*(1535)$ two curves are presented: The dashed line 
refers to the original set of input parameters with strong coupling (s) while 
the solid line refers to the new set with weaker coupling (w).
}
\label{width_fig}
\end{figure}

In Fig. \ref{width_fig} we show the energy dependence of the 
$N^* \rightarrow N\omega$ widths versus the off-shell masses $\mu$ of the 
resonances. Results were obtained with the $\omega$ spectral 
function of ref. \cite{krivo02}. 
We show two solutions for the $N^*(1535)$ resonance: The dashed line 
refers to the original set of input parameters with strong coupling (s) while 
the solid line refers to the new set with weaker coupling (w).

%%%%%%%%%%%%%%%%%%%%%%%%%%%%%%%%%%%%%%%%%%%%%%%%%%%%%%%%%%%%%%%%%%%%%%%%%
%%%%%%%%%%%%%%%%%%%%%%%%%%%%%%%%%%%%%%%%%%%%%%%%%%%%%%%%%%%%%%%%%%%%%%

When the off-shell mass of a $N^*$ resonance exceeds the $\omega$-meson production 
threshold given by the $\omega$ pole mass ($m_N + m_{\omega}$), the decay width 
sharply increases since the narrow $\omega$ peak acts similar as a $\delta$-function. 
The width of a resonance which lies below the $\omega$ pole and has 
already there a noticeable branching to this decay channel acquires a large width at 
the $\omega$ pole. Such a step-like behavior is not unusual. If there is a wide
potential barrier of height $U_0 ~~(= m_R + m_{\omega})$ the life time of a 
resonant state vanishes at an energy $E>U_0 ~~(\mu > m_R + m_{\omega})$. 
Moreover, the life time can vanish sharply if the potential barrier is flat 
(a square wall). Deeply below $U_0$, the barrier penetration factor 
$D(E)=exp(-2\sqrt{2m(U_0 - E})a)$ is almost constant 
(with $a$ being the barrier width), just below $U_0$ it increases 
exponentially, and just above $U_0$ $D(E)=1$.

\begin{figure}[h]
\begin{center}
\leavevmode
\epsfxsize = 8cm
\epsffile[60 50 430 410]{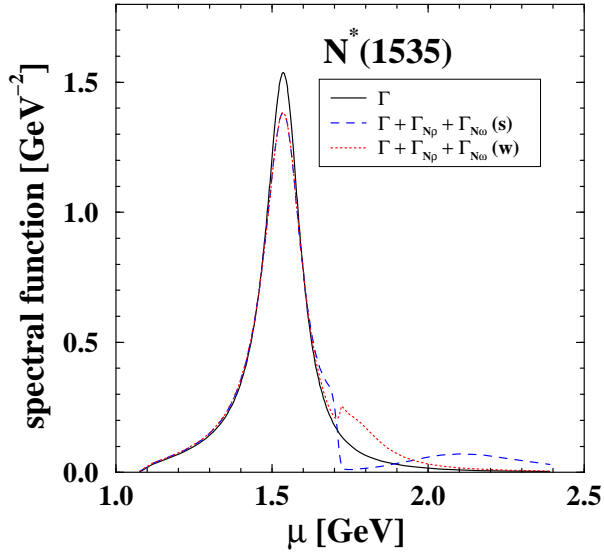}
\end{center}
\caption{Spectral distribution of the $N^*(1535)$ resonance using 
an energy dependent total width, and adding the partial 
$N\rho$ and $N\omega$ widths. The two cases of 
a strong (s) and weak (w) coupling to the $N\omega$ channel are 
distinguished.
}
\label{Nspec_fig}
\end{figure}

The $N^*(1535)$ off-shell partial width is especially large for the 
strong coupling amplitude (s), being 16 GeV above the $\omega$ threshold. 
We interpret this large $N^*(1535)$ width as an indication for a dissolving 
resonant state $N^*(1535)$ at running masses $\mu$ above the $\omega$ 
production threshold. Fig. \ref{Nspec_fig} shows the spectral function of 
the $N^*(1535)$. When large values of the decay width appear 
(dashed curve), also the resonance spectral function
drops sharply above the $\omega$ production threshold which demonstrates 
from another side the disappearance of the resonant states with increasing 
running mass. 
%%%%%%%%%%%%%%%%%%%%%%%%%%%%%%%%%%%%%%%%%%%%%%%%%%%%%%%%%%%%%%%%%%%%%%%%%
%%%%%%%%%%%%%%%%%%%%%%%%%%%%%%%%%%%%%%%%%%%%%%%%%%%%%%%%%%%%%%%%%%%%%%
The implications of a full dissolution of the $N^*(1535)$ are 
quite straightforward: Integrals over 
the running mass $\mu$ should be cut at $\mu > m_R + m_{\omega}$. 
In our model such an effect arises automatically through the appearance 
of the large total resonance width in the denominator of the 
Breit-Wigner distribution. The large $N^*(1535)$ width demonstrates 
that we are somewhere above a potential barrier where the resonance 
does not exist any more. However, in this case the resonance model 
allows to treat all resonances on the same footing, i.e. 
accounting in the calculations of the cross 
section for the full energy range $m_p + m_{\pi} < \mu < \sqrt s - m_p$. 
For the $N^*(1535)$ the interval $m_p + m_{\omega} < \mu < \sqrt s - m_p$ 
does not significantly influence observables due to its suppression 
through the large width. However, the scaling by the Blatt-Weisskopf factor 
leads to a suppression of $\Gamma^{R}_{N\omega}$ at $\mu \ge 2$ GeV 
which results in a small bump in the  $N^*(1535)$ 
spectral function at high values of $\mu$. This suppression gives some additional 
strength to the cross section at high energies. This fact has, however, 
no influence on the threshold behavior of the cross section 
on which the present work focuses. Relatively large $N\omega$ 
widths are not too surprising. According to the $SU(3)$ symmetry 
the $\omega$ coupling to nucleons is 3 times greater than the $\rho$ 
coupling. One can therefore expect that 
at equal kinematical conditions the off-shell $N\omega$ widths 
above the $\omega$ production threshold will typically be an order of 
magnitude greater than the $N\rho$ widths.
The off-shell partial widths of the weakly coupled $N^*(1535)$ (solid curve) 
and of the other resonances are below 1 GeV. 

In this context it should be also mentioned that 
expression (\ref{GM}) for the energy dependent $\omega$ width 
$\Gamma_\omega (M)$ contains still uncertainties away from the $\omega$ peak: 
First, the coupling constants $f_{\omega \rho \pi }$ and 
$f_{\rho \pi \pi }$ in eq. (\ref{GM}) can be $M$-dependent. 
Furthermore, the Blatt-Weisskopf factor which is, however, 
not included here would lead to a suppression of the off-shell widths 
above and enhancement below the resonance masses. This factor is usually 
applied to two-body decays. However, the $\omega$ decay has a 
three-body final state and thus an appropriate modification 
of the spectral function 
is not straightforward. Notice that the $\omega \rightarrow 3\pi$ decay 
cannot simply be treated as a quasi two-body decay either,
due to the coherent superposition of the $\rho$-meson propagators 
entering the integrand of Eq. (\ref{GM}).
In the case of the strong $N^*(1535)$ coupling  an enhancement 
of $\Gamma_\omega$ at $M<m_{\omega}$ would
result in a further increase of the off-shell part of the cross 
section as discussed below, but not in an increase of 
its measurable peak contribution. For the weak coupling, 
where the background is not very pronounced the cross section is 
practically not affected by this uncertainty.
%%%%%%%%%%%%%%%%%%%%%%%%%%%%%%%%%%%%%%%%%%%%%%%%%%%%%%%%%%%%%%%%%%%%%%%%%
\section{The $pp\rightarrow pp\omega$ cross section}
%%%%%%%%%%%%%%%%%%%%%%%%%%%%%%%%%%%%%%%%%%%%%%%%%%%%%%%%%%%%%%%%%%%%%%%%%
In this section we turn to the calculation of the $\omega$ production 
in $pp$ reactions. We distinguish between the two scenarios of a strong 
and a weak $N^*(1535)N\omega$ coupling. The decay modes 
of the other resonances are kept fixed as determined in \cite{krivo02} 
and given in Table 1. 
%%%%%%%%%%%%%%%%%%%%%%%%%%%%%%%%%%%%%%%%%%%%%%%%%%%%%%%%%%%%%%%%%%%%%%%%%
\subsection{Strong $N^*(1535)-N\omega$ coupling}
\begin{figure}[h]
\begin{center}
\leavevmode
\epsfxsize = 8cm
\epsffile[60 50 430 410]{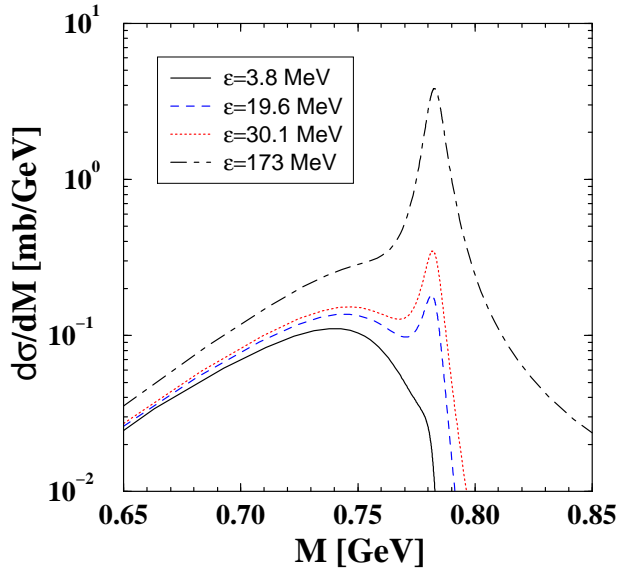}
\end{center}
\caption{Differential $pp\rightarrow pp\omega$ cross section as a 
function of the $\omega$ mass $M$ calculated at different values 
of the excess energy $\epsilon$. 
}
\label{sig0_fig}
\end{figure}

%%%%%%%%%%%%%%%%%%%%%%%%%%%%%%%%%%%%%%%%%%%%%%%%%%%%%%%%%%%%%%%%%%%%%%%%%
First we consider the differential cross section $d\sigma/dM$. 
Fig. \ref{sig0_fig} shows the differential cross section 
$d\sigma/dM$ for typical excess energies of the SATURNE 
($\epsilon = 3.8,~19.6,~30.1$ MeV) 
and the COSY-TOF ($\epsilon = 173$ MeV) experiments. At small excess energies 
the cross section is dominated by off-shell omega production with masses 
far below the $\omega$ pole mass of 782 MeV. At the lowest 
considered energy, i.e. 3.8 MeV, there is even no clear $\omega$ peak 
structure visible in the spectrum whereas at larger values of 
$\epsilon$ the $\omega$ peak is well developed. With increasing 
energy the off-shell production with $\omega$ masses far below 
the quasi-particle pole becomes less important. E.g. at 173 MeV 
the off-shell strength is already moderate. Here the cross section 
originates to most extent from the $\omega$ peak but   
the off-shell part leads still to a non-vanishing renormalization 
of the cross section. 
%%%%%%%%%%%%%%%%%%%%%%%%%%%%%%%%%%%%%%%%%%%%%%%%%%%%%%%%%%%%%%%%%%%%%%%%%
%%%%%%%%%%%%%%%%%%%%%%%%%%%%%%%%%%%%%%%%%%%%%%%%%%%%%%%%%%%%%%%%%%%%%%
%%%%%%%%%%%%%%%%%%%%%%%%%%%%%%%%%%%%%%%%%%%%%%%%%%%%%%%%%%%%%%%%%%%%%%%%%
\begin{figure}[h]
\begin{center}
\leavevmode
\epsfxsize = 8cm
\epsffile[60 50 430 410]{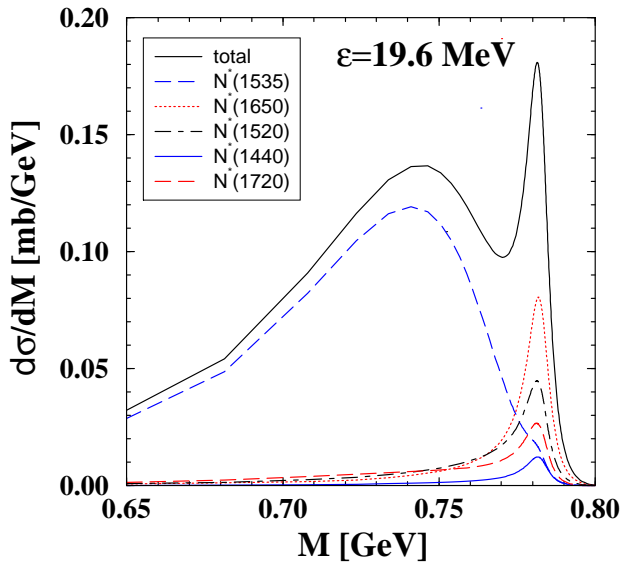}
\end{center}
\caption{Differential $pp\rightarrow pp\omega$ cross section as a 
function of the $\omega$ mass $M$ calculated at an excess 
energy of $\epsilon=19.6$ MeV. The contributions from the various resonances 
are shown separately.
}
\label{sig1_fig}
\end{figure}
%%%%%%%%%%%%%%%%%%%%%%%%%%%%%%%%%%%%%%%%%%%%%%%%%%%%%%%%%%%%%%%%%%%%%%
To explore the origin of the large off-shell 
component in more detail Fig. \ref{sig1_fig} shows 
separately the contributions 
of the various resonances at an excess energy of $\epsilon=19.6$ MeV.
As can be seen from there the decay of the $N^*(1535)$ is responsible 
for the large off-shell $\omega$ production with masses far below the 
pole mass whereas the off-shell contributions of the 
other resonances are small. However, at the $\omega$ pole the
situation is opposite, i.e. the contribution from the $N^*(1535)$ is 
strongly suppressed. The other resonances show a clear peak structure.

The reason for this peculiar behavior of the 
$N^*(1535)$ lies in its large coupling to the $N\omega$ channel already 
far below the threshold (given by the physical $\omega$ mass). Since the 
decay width is large at the resonance pole, i.e. 167 MeV below the 
$\omega$ pole, this resonance provides significant strength to the 
 $\omega$ production in a kinematical regime where the $\omega$ meson 
is far off-shell. When the $\omega$ mass approaches the on-shell value the 
partial decay width to the $N\omega$ channel increases
strongly. However, this strong increase suppresses the on-shell 
production of the meson. The suppression 
is already reflected in the spectral distribution of the  $N^*(1535)$ 
resonance, shown in Fig. \ref{Nspec_fig}. 
This is a general feature for the decay of a broad
resonance to another particle which arises under 
particular kinematical conditions. 
As discussed by Knoll for the case of the $\rho$ decay \cite{knoll} the 
opening of new decay channels suppresses generally the corresponding contributions 
at the on-shell values. The reason lies in the non-perturbative way by 
which the new channel enters into the resonance spectral function
where it adds to the total resonance width \cite{knoll}. In the case of the 
$N^*(1535)$ this effect is particularly pronounced due to the large 
$N\omega$ decay width and the corresponding kinematical conditions. 
At $\epsilon = 19.6$ MeV, the background generated by the 
$N^*(1535)$ stays below the experimental background \cite{hibou99}.
It is also smooth enough that it does not generate a visible structure
in the data \cite{hibou99}. Hence, the possible existence of 
such an off-shell background is not in contradiction to current 
experimental facts.

\begin{figure}[h]
\begin{center}
\leavevmode
\epsfxsize = 8.5cm
\epsffile[40 30 570 440]{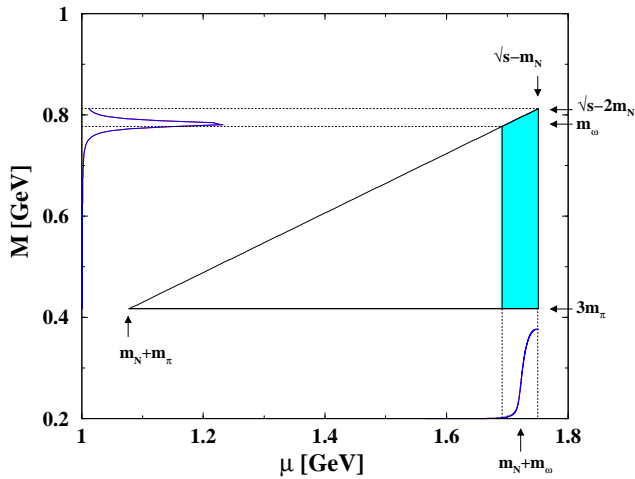}
\end{center}
\caption{The integration region for the $\omega$ production cross 
section is displayed. The integration is performed in the two-dimensional 
parameter space of the off-shell $\omega$-meson mass $M$ 
and the off-shell $N^*$ mass $\mu$. Along the horizontal axis the behavior
of the $N^* \rightarrow N\omega$ width as a function of $\mu$ is 
indicated schematically, on the vertical axis the $\omega$-meson 
spectral distribution is indicated.
The sharp increase of the $N^* \rightarrow N\omega$ width can result 
in a strong suppression of the integrand in the dashed region where 
the $\omega$ peak appears.
}
\label{integ_fig}
\end{figure}

The mechanism which is 
responsible for the occurrence of the strong off-shell $\omega$ 
contribution is illustrated in Fig. \ref{integ_fig}. 
We show the kinematical limits for the integrations 
of the cross section, Eq. (\ref{sigNNV}). The integration region is the 
two-dimensional parameter space of the off-shell $\omega$-meson mass $M$ 
and the off-shell $N^*$ mass $\mu$. Along the horizontal axis we 
plotted schematically the $N^* \rightarrow N\omega$ 
width as a function of $\mu$. 
On the vertical axis, we show schematically the Breit-Wigner function 
for the $\omega$-meson, which enters into Eq. (\ref{sigNNV}).
The sharp increase of the $N^* \rightarrow N\omega$ width can result 
in a strong suppression of the integrand in the shaded area 
due to Breit-Wigner function (\ref{BW}). 
The $\omega$ peak appears just in the region where
such a suppression is possible. The occurrence of the "theoretical background" 
has therefore a simple kinematical origin. Whether it occurs depends on 
the off-shell $N^* \rightarrow N\omega$ width in the 
vicinity of the $\omega$-meson production threshold. 
The integration boundaries for this example are determined 
for an excess energy of $\epsilon$ = 30 MeV. Notice that the 
suppression of the $\omega$ peak in the strong coupling regime 
is still present at increasing $\sqrt s$. 
%%%%%%%%%%%%%%%%%%%%%%%%%%%%%%%%%%%%%%%%%%%%%%%%%%%%%%%%%%%%%%%%%%%%%%%%%
%%%%%%%%%%%%%%%%%%%%%%%%%%%%%%%%%%%%%%%%%%%%%%%%%%%%%%%%%%%%%%%%%%%%%%
%%%%%%%%%%%%%%%%%%%%%%%%%%%%%%%%%%%%%%%%%%%%%%%%%%%%%%%%%%%%%%%%%%%%%%%%%
\begin{figure}[h]
\begin{center}
\leavevmode
\epsfxsize = 8cm
\epsffile[60 50 430 410]{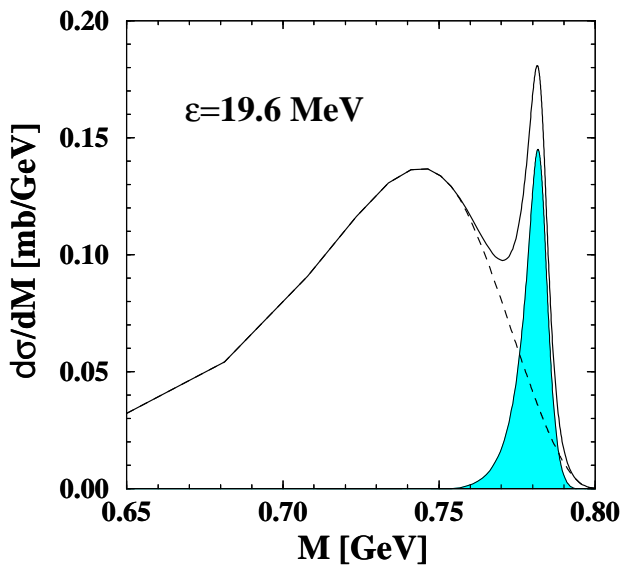}
\end{center}
\caption{Differential $pp\rightarrow pp\omega$ cross section at 
an excess energy of $\epsilon=19.6$ MeV. The theoretical background 
due to off-shell $\omega$ production (dashed line) is subtracted from 
the full cross section (solid line). The hatched area shows the 
remaining contribution from the $\omega$ peak.  
}
\label{back_fig}
\end{figure}
%%%%%%%%%%%%%%%%%%%%%%%%%%%%%%%%%%%%%%%%%%%%%%%%%%%%%%%%%%%%%%%%%%%%%%
Although the integration 
region dispayed in Fig. 6 increases the effective 
integration region is restricted by the $\omega$ pole mass. 
Distinct from other resonances, the contribution of the $N^*(1535)$ 
to the cross section does therefore not increase with energy. 
This explains why the theoretical "background" becomes less important 
at high energy.

\begin{figure}[h]
\begin{center}
\leavevmode
\epsfxsize = 8cm
\epsffile[60 50 430 410]{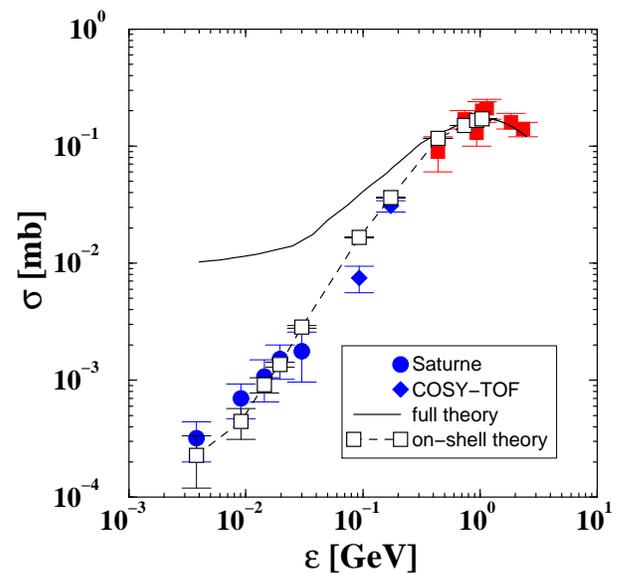}
\end{center}
\caption{Exclusive $pp\rightarrow pp\omega$ cross section 
obtained in the resonance 
model. Data are taken from \protect\cite{hibou99,cosy01} and 
\protect\cite{disto01,flaminio}. The solid curve shows the full cross 
section including the contributions from off-shell $\omega$ production. 
The dashed curve corresponds to the renormalized cross section where 
only the experimentally detectable contribution from the $\omega$ 
peak is taken into account. 
}
\label{sig2_fig}
\end{figure}

At small excess energies the cross section receives its major
strength from off-shell $\omega$ production. However, experimentally 
only the $\omega$ peak can be identified in missing mass spectra and 
the off-shell part of the cross section is attributed 
to the general experimental background. Hence, for a meaningful 
comparison to data the same procedure has to be applied to theory which 
means to take only the contribution form the $\omega$ peak into
account. The subtraction of this 
theoretical ``background'' from off-shell meson production is 
demonstrated in Fig. \ref{back_fig} for $\epsilon=19.6$ MeV. 
Like in the experimental analysis \cite{cosy01} the background is 
smoothly interpolated by splines. The remaining peak contribution 
resembles a Breit-Wigner distribution around the $\omega$ pole mass 
which, due to the energy dependence of $\Gamma_\omega$ and the phase space
factor $\Phi_3(\sqrt s,m_p,m_p,M)$ entering into the cross section, is asymmetric. 
As soon as a clear peak structure appears the 
decomposition of the cross section is straightforward and rather unique 
as indicated by the error bars shown in Fig. \ref{sig2_fig}. 
%%%%%%%%%%%%%%%%%%%%%%%%%%%%%%%%%%%%%%%%%%%%%%%%%%%%%%%%%%%%%%%%%%%%%%%%%
%%%%%%%%%%%%%%%%%%%%%%%%%%%%%%%%%%%%%%%%%%%%%%%%%%%%%%%%%%%%%%%%%%%%%%
The correspondingly renormalized total cross section is shown in Fig. 
\ref{sig2_fig} as a function of the excess energy $\epsilon$. 
The theoretical calculation provides 
now a very accurate description of the exclusive experimental $\omega$ 
cross section from close to threshold 
\cite{hibou99,cosy01} up to excess energies of several GeV. 
In particular, at low energies the slope of the cross section 
is well reproduced. The full theoretical cross section, i.e. 
including off-shell $\omega$ 
production, is also shown in Fig. \ref{sig2_fig}. The total 
cross section is systematically larger than the renormalized and 
the experimental cross sections. Above excess energies of the DISTO 
point \cite{disto01} (440 MeV) the off-shell production starts to become 
negligible and both curves almost coincide. Due to the increasing phase space 
of the $\omega$ width at large excess energies there arise new off-shell 
contributions from the high momentum tail of the $\omega$ spectral 
function. Since the present work concentrates on the threshold behavior, 
we accounted only partially for high momentum corrections to the 
cross section which are small anyhow. However, close to threshold 
the contribution from the off-shell 
``background'' to the total cross section is about one order of magnitude 
larger than the measurable pole part of the cross section. 
%%%%%%%%%%%%%%%%%%%%%%%%%%%%%%%%%%%%%%%%%%%%%%%%%%%%%%%%%%%%%%%%%%%%%%%%%
\subsection{Weak $N^*(1535)-N\omega$ coupling}
%%%%%%%%%%%%%%%%%%%%%%%%%%%%%%%%%%%%%%%%%%%%%%%%%%%%%%%%%%%%%%%%%%%%%%%%%
If one assumes a coupling of the  $N^*(1535)$ to the $N\omega$ channel 
which is 6 to 8 times weaker than in the previous case the 
$pp\rightarrow pp\omega$ cross section has a completely 
different behavior close to threshold. This scenario is based 
on the alternatively selected data set for this resonance (see the
Appendix). We show the differential cross 
section again for $\epsilon=19.6$ MeV. First of all, the large
contribution form off-shell $\omega$ production vanishes. There remains 
still strength at masses far below the $\omega$ pole but compared to
the previous case the off-shell production is almost negligible. 
On the other hand, in contrast to the case of a strong 
$N^*(1535)-N\omega$ coupling dicussed in the previous section, 
the on-shell $\omega$ production from the $N^*(1535)$ is not 
suppressed but contributes maximally to the on-shell cross section. 
Since the contributions from the other resonances remain unchanged
the magnitude of the cross section at the $\omega$ pole mass is about 
30 \% larger as in the previous case. The deviation increases with 
decreasing $\sqrt s$.

Now the ``theoretical background'' due to off-shell 
production is much smaller than in the previous case 
and thus the distinction between the on-shell and the moderate off-shell 
part of the cross section is not straightforward. 
Nevertheless, very close to threshold 
the background is present and the cross section
does not vanish when the excess energy $\epsilon$ goes to zero. 
When the $\omega$ meson is treated as an elementary field \cite{Nak98} 
the cross section, according to phase space, vanishes by construction
at the threshold. However, here the physical threshold 
for the $\omega$ production is given by the $3 m_\pi$ threshold 
instead of the $\omega$ pole mass. 
The threshold behavior is qualitatively similar to a recent 
calculation from the J\"ulich group \cite{Nak00} where the finite 
mass distribution of the $\omega$ has been taken into account. 
However, close to threshold our result is significantly larger 
and overestimates the SATURNE \cite{hibou99} and the COSY \cite{cosy01} 
data. A subtraction of the small and not so well 
defined off-shell background is now a delicate procedure which depends 
strongly on the particular experimental treatment of the background. 
We estimated the magnitude of this effect and found that it 
might bring the theoretical result into slightly 
better agreement with experiment at the lowest measured 
excess energies \cite{hibou99}. But already at $\epsilon$ = 30 MeV and 
in particular for the COSY points \cite{cosy01} such corrections are 
practically zero and thus the experimental results are still 
overestimated. In the COSY regime  the tendency is similar 
as in the calculations from \cite{Nak98,Nak00} which show 
also a too steep increase with energy.
%%%%%%%%%%%%%%%%%%%%%%%%%%%%%%%%%%%%%%%%%%%%%%%%%%%%%%%%%%%%%%%%%%%%%%%%%
\begin{figure}[h]
\begin{center}
\leavevmode
\epsfxsize = 8cm
\epsffile[60 50 430 410]{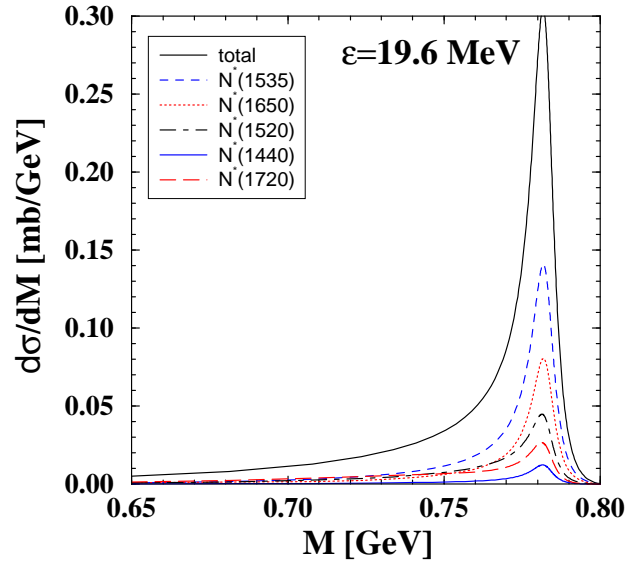}
\end{center}
\caption{Differential $pp\rightarrow pp\omega$ cross section as a 
function of the $\omega$ mass $M$ calculated at an excess 
energy of $\epsilon=19.6$ MeV. The contributions from the various resonances 
are shown separately.
}
\label{sig1b_fig}
\end{figure}
%%%%%%%%%%%%%%%%%%%%%%%%%%%%%%%%%%%%%%%%%%%%%%%%%%%%%%%%%%%%%%%%%%%%%%
At large excess energies above 1 GeV the cross section is somewhat 
reduced compared to the case with large $N^*(1535)-N\omega$ coupling 
and also compared to experiment. The reason is missing strength from 
this resonance at high energies. This behavior is already reflected in the
spectral function, Fig. \ref{Nspec_fig}. The large $N^*(1535)-N\omega$ 
coupling depletes the spectral function at the $\omega$ threshold and 
shifts strength to the high energy tail which is missing in the second 
case. There the strength is concentrated around and a few hundred MeV above
the $\omega$ threshold which results in the enhanced cross section 
in the COSY regime. 

In ref. \cite{disto}, the experimental results were compared 
to a cross section which is assumed to be proportional to the
$NN\omega$ phase space corrected by the proton-proton $s$-wave 
FSI and the finite $\omega$ width. It was pointed out that
the energy of the protons is sufficiently high to excite higher partial 
waves where the proton-proton correlations are not important. 
A smooth interpolation to lower $\sqrt s$ should then be possible 
if there exists no strong coupling of nucleon resonances 
to the $N\omega$ channel. Apparently, our model is based on the 
opposite assumption, namely that the dynamics is 
governed by nucleon resonances. With respect to the FSI, 
our model is complementary in the sense that we account for 
the $N\omega$ FSI through the inclusion of resonances 
whereas the proton-proton FSI is missing. Near the threshold the
proton-proton FSI generally enhances the amplitude which is relevant for the 
SATURNE data \cite{hibou99}. For the weak $N^*(1535)$ coupling 
the amplitude is in our model already overestimated already due to the 
inclusion of the $N\omega$ FSI. The reproduction of the data 
with a $NN\omega$ contact term for strongly correlated $s$-wave nucleons 
would require an amplitude of opposite sign and a delicate cancellation 
between the proton-proton and the $N\omega$ FSI at threshold. 
The resonance model proposed by Teis et al. \cite{Teis} contains such a
contact term for a direct $NN\pi$ coupling which could be attributed to the 
$NN$ FSI and/or a non-resonant background. Its contribution is, however, not so important 
at threshold for the $pp\pi$ final states. 
This fact can be interpreted to mean that the proton-proton FSI and/or a background
are effectively already included into the phenomenological matrix elements 
for the resonance production. 
%%%%%%%%%%%%%%%%%%%%%%%%%%%%%%%%%%%%%%%%%%%%%%%%%%%%%%%%%%%%%%%%%%%%%%%%%
\begin{figure}[h]
\begin{center}
\leavevmode
\epsfxsize = 8cm
\epsffile[60 50 430 410]{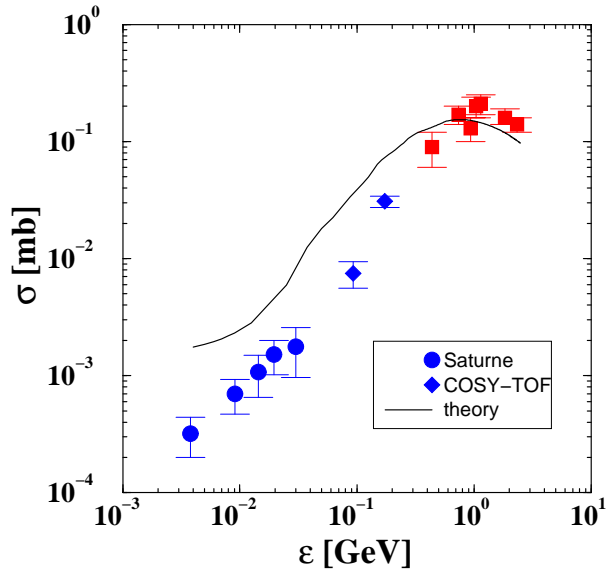}
\end{center}
\caption{Exclusive $pp\rightarrow pp\omega$ cross section 
obtained in the resonance 
model. Data are taken from \protect\cite{hibou99,cosy01} and 
\protect\cite{disto01,flaminio}. 
}
\label{sig2b_fig}
\end{figure}
%%%%%%%%%%%%%%%%%%%%%%%%%%%%%%%%%%%%%%%%%%%%%%%%%%%%%%%%%%%%%%%%%%%%%%
In summary, the comparison to the data favors the 
calculation based on the large 
$N^*(1535)N\omega$ branching ratio. An additional fact which supports 
this picture is the strongly anisotropic $pp\omega$ angular distribution 
observed in \cite{cosy01} at $\epsilon=173$ MeV. As can be seen from 
Table 1 only the $N^*(1535)$ has a large $s$-wave component in the $N\omega$ 
decay. The contributions from the other resonances involve higher 
partial waves which lead to anisotropic angular distributions.  
As we have seen, the strong $N^*(1535)N\omega$ coupling 
suppresses the contribution from this resonance at the $\omega$ 
pole which leads to the following 
decomposition of the on-shell cross section at $M=m_\omega$: 
$N^*(1535)$ (7\%), $N^*(1650)$ (26\%), $N^*(1520)$ (17\%), 
$N^*(1440)$ (22\%) and $N^*(1720)$ (26\%). Hence the on-shell cross 
section contains only a small $s$-wave component. In the alternative 
case the $s$-wave contribution from the $N^*(1535)$ is much larger, 
the cross section decomposes according to: 
$N^*(1535)$ (35\%), $N^*(1650)$ (18\%), $N^*(1520)$ (12\%), 
$N^*(1440)$ (15\%) and $N^*(1720)$ (18\%). 
An highly anisotropic angular distribution indicates furthermore 
that contribution from meson exchange currents which are missing 
in our treatment are small \cite{Nak98}. 

%%%%%%%%%%%%%%%%%%%%%%%%%%%%%%%%%%%%%%%%%%%%%%%%%%%%%%%%%%%%%%%%%%%%%%%%%
\section{Summary}
%%%%%%%%%%%%%%%%%%%%%%%%%%%%%%%%%%%%%%%%%%%%%%%%%%%%%%%%%%%%%%%%%%%%%%%%%
The $\omega$ production in nucleon-nucleon collisions has been 
described via the excitation of nucleon resonances. 
The approach is based on the 
extended VMD (eVMD) model which successfully describes the mesonic 
$R\rightarrow NV$ and radiative $R\rightarrow N\gamma$ decays of 
nucleon resonances. Among the considered resonances the $N^*(1535)$ 
turns out to play a special role for the $\omega$ production. The 
reason is a large decay mode to the $N\omega$ channel in a kinematical 
regime where the $\omega$ is far off-shell. A strong
$N^*(1535)N\omega$ coupling is implied 
by the available electro- and photoproduction data. 
As a consequence large off-shell contributions in the $\omega$ production 
cross section appear. In particular close to threshold the 
off-shell production is dominant. On the other hand, the on-shell 
production of $\omega$ mesons at their physical masses is suppressed, since
above the $\omega$ production threshold the $N^*(1535)$ acquires
a large width and dissolves. 
This feature appears generally when a broad resonance has a 
large branching to a narrow state already far below the pole mass of 
the produced particle. In quantum mechanics, a similar behavior is
experienced by resonance states at energies just above a potential barrier.

We found that near threshold, the off-shell production of the $\omega$'s
becomes dominant. This part of the cross section can, 
however, be hardly identified experimentally and is attributed currently 
to the background. To compare to data we applied the same procedure as 
experimentalists: The theoretical "background" from the off-shell 
production was subtracted and only the measurable pole part 
of the cross section was taken into account. Doing so, the 
available data are accurately reproduced starting from energies very 
close to threshold up to energies significantly above threshold without 
adjusting any new parameters. At small 
excess energies the full cross section is about one order of magnitude 
larger than the measurable pole part. 

Since the observed results depend crucially on the role of the $N^*(1535)$ 
we considered also an alternative scenario which is possible 
within experimental uncertainties. Since the $N\omega$ decay of this 
resonance has not been measured directly, the existing $N\rho$ data 
leave some freedom to fix the eVMD model parameters. A different 
normalization to the $N\rho$ channel, making thereby use of an 
alternative set of quark model predictions, allows to reduce the 
$N\omega$ decay mode by maximally a factor of 6 to 8, however, at 
the expense of a slightly worse reproduction of the existing data set. 
With the reduced $N\omega$ coupling the $N^*(1535)$ shows no 
extraordinary behavior and the off-shell contributions 
are substantially reduced. However, this resonance contributes 
now fully to the peak part of the cross section which 
leads to a significant overestimation of the experimental data around and 
several 100 MeV above threshold. The discrepancy is quite strong 
and can apparently not be attributed solely to the $NN$ FSI and/or a non-resonant 
background.

We conclude that, consistent with the analysis of electro- 
and photoproduction data, the measured $\omega$ production in 
$pp$ reactions favors the scenario of a large coupling of the 
 $N^*(1535)$ to the $N\omega$ channel. The consequences are large 
off-shell contributions in the $\omega$ production cross section 
around threshold.  Experimentally this part of the cross section 
is hardly accessible since it is hidden in the general experimental 
background. However, if the off-shell $\omega$ production is large, 
the number of dileptons produced in $pp$ collisions should 
be significantly greater than one would expect 
from estimates based on the on-shell production of $\omega$'s. 
This effect should manifest itself in the description of 
the dilepton production in $pp$ collisions at kinetic beam energies 
which correspond to the $\omega$-meson threshold. It can have 
consequences for the dilepton production in heavy-ion collisions.

The hypothesis of a large off-shell $\omega$ production is also 
appealing from another point of view. The $\phi$ meson decays dominantly 
to the $K{\bar K}$ mode with a threshold very close to the physical 
$\phi$ mass. Hence one does not expect sizable off-shell effects in this 
case. When off-shell $\omega$'s are counted, the $\phi/\omega$ ratio 
could come into agreement with the OZI rule, i.e. the experimentally 
observed deviations from the OZI predictions for the $\phi/\omega$ ratio 
might be explained by the neglection of the off-shell $\omega$-meson 
production. 

\subsection*{Acknowledgments}
The work was supported by GSI (Darmstadt) under the contract 
T\"{U}F\"{A}ST, by the Plesler Foundation, and by the Deutsche
Forschungsgemeinschaft under contract No.~436RUS113/367/0(R).
%%%%%%%%%%%%%%%%%%%%%%%%%%%%%%%%%%%%%%%%%%%%%%%%%%%%%%%%%%%%%%%%%%%%%%%%%
\section{Appendix}
%%%%%%%%%%%%%%%%%%%%%%%%%%%%%%%%%%%%%%%%%%%%%%%%%%%%%%%%%%%%%%%%%%%%%%%%%
The eVMD model parameters for the description of $N\omega$ decays 
of nuclear resonances are determined form experimental data. 
%%%%%%%%%%%%%%%%%%%%%%%%%%%%%%%%%%%%%%%%%%%%%%%%%%%%%%%%%%%%%%%%%%%%%%%%%
\begin{figure}[h]
\begin{center}
\leavevmode
\epsfxsize = 8.5cm
\epsffile[40 40 390 690]{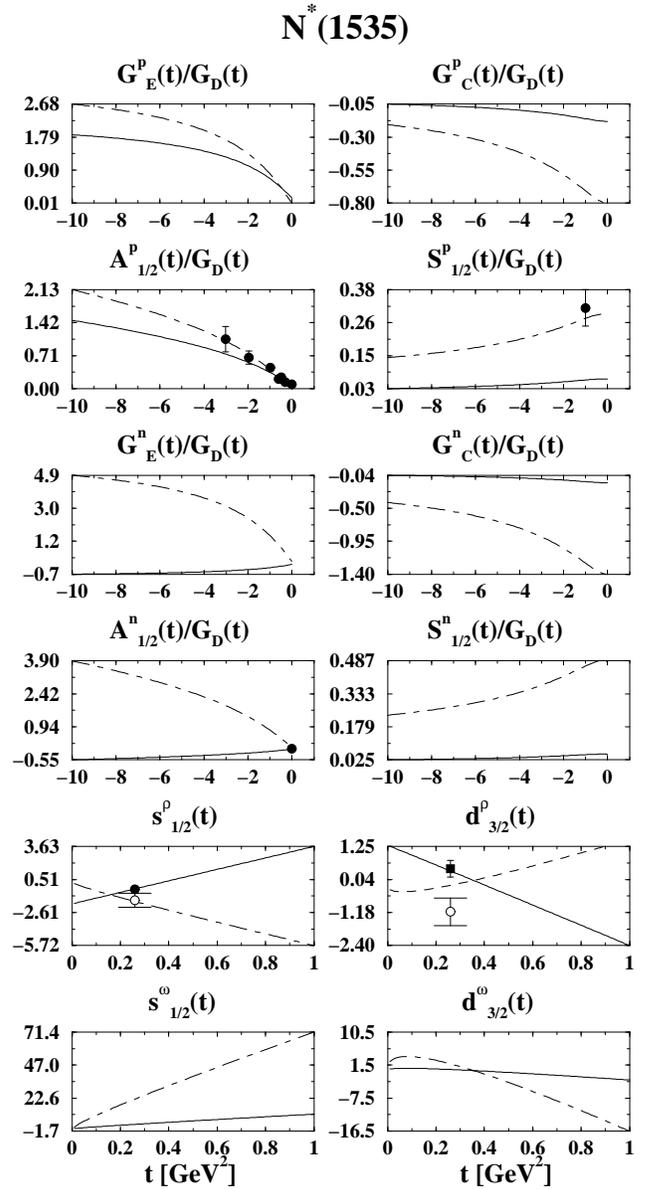}
\end{center}
\caption{Electric and Coulomb transition form factors $G_E$ and $G_C$ (in GeV$^{-1}$), 
helicity amplitudes $A_{1/2}$ and $S_{1/2}$ (in GeV$^{-\frac{1}{2}}$), 
and partial-wave amplitudes (in GeV$^{-2}$) are shown for the $N^*(1535)$ 
decay to the $\rho $- and $\omega$-meson channels. 
The value $G_{D}(t) = 1/(1 - t/0.71)^2$ is the 
dipole function. The experimental photo- and
electro-production data $A_{1/2}$ and $S_{1/2}$ for the $p^*(1535)$ 
are taken from \protect\cite{Groom:2000in}.
The $\rho$ meson decay amplitudes (filled symbols) are from 
\protect\cite{Koniuk:1982ej,Manley:1992yb,Groom:2000in}, the opaque circles are 
the quark model predictions from \protect\cite{Capstick:1994kb}. The dot-dashed lines
correspond to fit to the strong coupling set 
\protect\cite{Koniuk:1982ej,Manley:1992yb,Groom:2000in} while the solid lines
correspond to the weak coupling set \protect\cite{Capstick:1994kb}.
}
\label{Nfit1_fig}
\end{figure}
%%%%%%%%%%%%%%%%%%%%%%%%%%%%%%%%%%%%%%%%%%%%%%%%%%%%%%%%%%%%%%%%%%%%%
Fig. 11 shows two different fits
to the available electro- and photoproduction data, $\pi N$ scattering multichannel 
analyses and quark model predictions for the $N^*(1535)$
resonance. The difference between the two procedure lies in 
the normalizations to the $\rho$-meson decay amplitudes. The original 
fit of \cite{krivo02} (dot-dashed lines) is based on the 
the results of refs. \cite{Koniuk:1982ej,Manley:1992yb,Groom:2000in}. The 
quark model of Koniuk \cite{Koniuk:1982ej}, 
the $\pi N$ analysis of Manley and Saleski \cite{Manley:1992yb}, and
the PDG \cite{Groom:2000in} give close predictions for the $s_{1/2}$ wave 
and predict the same sign for the $d_{3/2}$ wave. 
Using this set of data the solution is rather stable and yields 
a large $N\omega$ $s_{1/2}$ wave amplitude. 

The second set fit (solid lines) is based on the 
$s_{1/2}$ and $d_{3/2}$ $N\rho$ amplitudes taken from 
quark model calculations by Capstick and Roberts \cite{Capstick:1994kb}. 
Their values are significantly smaller than those proposed by 
\cite{Koniuk:1982ej,Manley:1992yb,Groom:2000in}. 
This allows a reduction of the $N\omega$ $s_{1/2}$ amplitude by 
a factor of 6 to 8, however, by the price of a moderately higher $\chi^2$. 
A further going reduction of the amplitude is hardly possible. 
For the second fit the deviation from the experimental $p^*(1535)$ Coulomb 
amplitude is larger and the reproduction of the transversal $p^*(1535)$ 
amplitude is also worse but the $\rho$-meson amplitudes are
better reproduced than with the first set of input parameters. 
Notice that the $d_{3/2}$ amplitude of the $\omega$-meson was set 
equal to zero to ensure an unique eVMD solution for the fitting procedure. 

%%%%%%%%%%%%%%%%%%%%%%%%%%%%%%%%%%%%%%%%%%%%%%%%%%%%%%%%%%%%%%%%%%%%%%%%%
%                                                                       %
%   BEGIN OF BIBLIOGRAPHY                                               %
%                                                                       %
%%%%%%%%%%%%%%%%%%%%%%%%%%%%%%%%%%%%%%%%%%%%%%%%%%%%%%%%%%%%%%%%%%%%%%%%%

%%%%%%%%%%%%%%%%%%%%%%%%%%%%%%%%%%%%%%%%%%%%%%%%%%%%%%%%%%%%%%

%%%%%%%%%%%%%%%%%%%%%%%%%%%%%%%%%%%%%%%%%%%%%%%%%%%%%%%%%%%%%%%%%%%%%%
\onecolumn[\hsize\textwidth\columnwidth\hsize
           \csname @onecolumnfalse\endcsname]
%%%%%%%%%%%%%%%%%%%%%%%%%%%%%%%%%%%%%%%%%%%%%%%%%%%%%%%%%%%%%%%%%%%%%%
\begin{table}
\begin{center}
\caption{Resonances ($R$) included into the $NN \rightarrow NN\omega$ cross section
trough the two-step mechanism $NN \rightarrow NR$, $R \rightarrow N\omega$.
The second column shows the total widths of the resonances. 
The third column shows the partial widths for $N\omega$ decays 
($\protect\sqrt{\Gamma_{N\omega}}$ in MeV$^{1/2}$). The next three columns show 
the partial-wave decomposition of the $N\omega$ widths, 
including signs of the amplitudes. The coupling 
constants $g_{M}$, $g_{E}$, and $g_{C}$
at the $\omega$ pole are given in the last three columns in 
units GeV$^{-1}$ for spin $J = 1/2$ and GeV$^{-l + 1}$ for $J = l + 1/2$. 
The $N\omega$ 
decay modes are determined by fitting the photo- and electro-production
data, results of the multichannel $\pi N$ partial-wave analyses and
quark model predictions.
In the case of the $N^*(1535)$, the results from the two different fits 
with a strong (s), respectively, and a weak (w) $N\omega$ branching are given.
}
\begin{tabular}{cccccccccc}
 Resonance & $ \Gamma_{0}$ [MeV] & &$\sqrt{\Gamma_{N\omega}}$ [MeV$^{1/2}$] 
& $N\omega$ & $N\omega$ & $N\omega$ & $g_{M}$ & $g_{E}$ & $g_{C}$\\
\hline
          &     &     &                         & $s_{1/2}$ & $d_{3/2}$ & \\
$N^*(1535)\frac{1}{2}^-$   & 150  & s:& 1.43    & 1.43   & 0.05  &      &        &       -28.03&       -42.67\\
                           &      & w:& 0.21    & 0.21   & 0.01  &      &        &       -4.22&       -6.34\\
$N^*(1650)\frac{1}{2}^-$   & 150  &   &0.97     & -0.97  & -0.02 &      &        &        2.01&        4.14\\
\hline
          &     &     &                          &$d_{1/2}$ & $d_{3/2}$ & $s_{3/2}$  \\
$N^*(1520)\frac{3}{2}^-$   & 120  &   & 0.29  & -0.02 &   0.03   & 0.28   & -7.67&       18.16&       46.13\\
\hline
          &     &     &                          &$d_{1/2}$ & $d_{3/2}$ & $g_{3/2}$  \\
$N^*(1675)\frac{5}{2}^-$   & 150  &   & 0.06  & 0.06 &   $<$ 0.01   & $<$ 0.01   &        2.07&       -1.61&      -10.50\\
\hline
          &     &     &                          &$p_{1/2}$ & $p_{3/2}$ &   \\
$N^*(1440)\frac{1}{2}^+$   & 350 &   &  $<$ 0.01  & $<$ 0.01 & $<$ 0.01 &  &      -14.14&        &       63.13\\
\hline
          &     &     &                          &$p_{1/2}$ & $p_{3/2}$ & $f_{3/2}$  \\
$N^*(1720)\frac{3}{2}^+$   & 150 &   &  5.69  & 5.29 & -2.09 & 0.14 &        0.14&       -8.27&      -37.73\\
\hline
          &     &     &                          &$f_{1/2}$ & $f_{3/2}$ & $p_{3/2}$  \\
$N^*(1680)\frac{5}{2}^+$   & 130 &   &  0.71  & 0.09 & 0.58 & 0.40 &       -1.34&      -11.75&       -7.98\\
\end{tabular}
\label{iso_tab1}
\end{center}
\end{table}
%%%%%%%%%%%%%%%%%%%%%%%%%%%%%%%%%%%%%%%%%%%%%%%%%%%%%%%%%%%%%%
\end{document}